\documentclass[a4paper,11pt]{article}
\pdfoutput=1
\usepackage{jheppub}

\usepackage{amsmath,bm,amssymb,amsthm,mathrsfs,mathtools}
\usepackage{subcaption,color}
\usepackage{ragged2e}
\usepackage{caption}
\usepackage{array}
\justifying\let\raggedright\justifying

\allowdisplaybreaks[4]
\usepackage{tikz}
\usepackage{tikz-feynman}
\tikzset{>=latex} 
\usepackage{multirow}
\usepackage{comment}
\usetikzlibrary{backgrounds}
\usetikzlibrary{shapes,matrix,trees}
\usetikzlibrary{arrows.meta}
\usepackage{centernot}
\usetikzlibrary{positioning}			
\usetikzlibrary{calc,through}			
\usetikzlibrary{decorations.pathreplacing}     
\usepackage{pgffor}                                        
\usetikzlibrary{decorations.pathmorphing}	
\usetikzlibrary{decorations.markings}
\tikzset{
	vector/.style={decorate, decoration={snake}, draw},
	provector/.style={decorate, decoration={snake,amplitude=2.5pt}, draw},
	antivector/.style={decorate, decoration={snake,amplitude=-2.5pt}, draw},
	fermion/.style={draw=black, postaction={decorate},
		decoration={markings,mark=at position .55 with {\arrow[draw=black]{>}}}},
	fermionbar/.style={draw=black, postaction={decorate},
		decoration={markings,mark=at position .55 with {\arrow[draw=black]{<}}}},
	fermionnoarrow/.style={draw=black},
	gluon/.style={decorate, draw=black,
		decoration={coil,amplitude=4pt, segment length=5pt}},
	scalar/.style={dashed,draw=black, postaction={decorate},
		decoration={markings,mark=at position .55 with {\arrow[draw=black]{>}}}},
	scalarbar/.style={dashed,draw=black, postaction={decorate},
		decoration={markings,mark=at position .55 with {\arrow[draw=black]{<}}}},
	scalarnoarrow/.style={dashed,draw=black},
	electron/.style={draw=black, postaction={decorate},
		decoration={markings,mark=at position .55 with {\arrow[draw=black]{>}}}},
	bigvector/.style={decorate, decoration={snake,amplitude=4pt}, draw},
	photon/.style={decorate, draw=black,decoration={snake,amplitude=4pt, segment length=5pt} }
}

\definecolor{ccblue}{rgb}{0.0,0.4,0.8}
\usepackage[]{hyperref}
\hypersetup{colorlinks=true,
	linkcolor=ccblue,
	urlcolor=ccblue,
	citecolor=ccblue}
\usepackage{amsmath}
\usepackage{amsfonts}
\usepackage{amssymb}
\usepackage{bbm}
\usepackage{booktabs}
\usepackage{slashed}
\usepackage[]{hyperref}
\usepackage{autobreak}
\definecolor{boxcolor}{RGB}{235,245,255}
\newcommand{\mybox}[1]{\begin{center}\fcolorbox{black}{boxcolor}{\parbox[c]{16.5cm}{#1}}\end{center}}
\newcommand{\inbox}[2]{\begin{tabular}{rl}{\tt In[#1]:= } #2\end{tabular}}
\newcommand{\out}[2]{\begin{tabular}{rl}{\tt Out[#1]:= } #2\end{tabular}}
\newcommand{\outbox}[2]{\begin{tabular}{rl}{Prints $\Rightarrow$\hspace{0.85mm}} #2\end{tabular}}
\newcommand{\boxsplit}{\rule{16.5cm}{0.5pt}}
\newcommand{\gibspace}[2]{\begin{tabular}{rl}{\tt \hphantom{aaaaaaa} } #2\end{tabular}}

\preprint{~ZU-TH-87/25}

\title{\boldmath Multiple Mellin-Barnes integrals with polygamma functions
}
\author[a,b]{Sumit Banik,}

\author[c,d]{Samuel Friot}

\emailAdd{sumit.banik@physik.uzh.ch}
\emailAdd{samuel.friot@universite-paris-saclay.fr}

\affiliation[a]{Physik-Institut, Universit\"at Z\"urich, Winterthurerstrasse 190, 8057 Z\"urich, Switzerland}
\affiliation[b]{PSI Center for Neutron and Muon Sciences, 5232 Villigen PSI, Switzerland}
\affiliation[c]{Universit\'e Paris-Saclay, CNRS/IN2P3, IJCLab, 91405 Orsay, France}
\affiliation[d]{Univ Lyon, Univ Claude Bernard Lyon 1, CNRS/IN2P3,\\
 IP2I Lyon, UMR 5822, F-69622, Villeurbanne, France}

\abstract{
Mellin-Barnes (MB) integrals appear in various branches of physics and mathematics and are, in particular, used as a standard tool for evaluating multi-loop, multi-scale Feynman integrals both analytically and numerically. Recent geometric approaches based on conic hulls and triangulations provide a systematic framework for computing multiple MB integrals in terms of multivariate series. These approaches have so far been limited to MB integrals whose integrands are ratios of products of Euler's gamma functions only. However, in Feynman integral calculus, MB integrals with polygamma functions naturally arise, for instance, after resolving singularities in the dimensional-regularisation parameter $\epsilon$ and expanding the MB integrand in powers of $\epsilon$, as done by the public codes \texttt{MB.m} and \texttt{MBresolve.m}.
In this paper, we extend the conic hull and triangulation methods to the computation of MB integrals having polygamma functions in their integrand. We show that 
the arguments of
polygamma functions can be treated in a similar way to the arguments of gamma functions when applying the conic hull and triangulation techniques to identify poles that would contribute to different series solutions. However, since the singularity structure of the polygamma function is different from that of the gamma function, we propose two different ways to compute MB integrals involving polygamma functions, depending on whether the MB integral has straight or non-straight contours. We have implemented these algorithms in an updated version of the \textit{Mathematica} package \texttt{MBConicHulls.wl}, and we illustrate their use with a set of examples from Feynman integral calculus.
}
\newpage

\begin{document} 
\maketitle

\section{Introduction}

Mellin-Barnes (MB) integrals are a powerful computational tool frequently used in various domains of physics, such as particle physics to compute multi-loop Feynman integrals~\cite{Smirnov:2012gma, Dubovyk:2022obc}, 
study of electromagnetic wave propagation in turbulence~\cite{Sasiela}, detector physics~\cite{Friot:2014ufa}, condensed matter physics~\cite{PhysRevB.101.235162}, etc. In mathematics, MB integrals are a useful tool for the theory of multivariate hypergeometric functions~\cite{KdF,Exton,Marichev,srivastava1985multiple}, in asymptotics \cite{Paris&Kaminsky}, to compute definite integrals~\cite{Gonzalez:2021vqh} and to study GKZ partial differential equations~\cite{Feng:2019bdx}, among others. MB integrals can even be found in economics in the study of option pricing~\cite{Aguilar2017SeriesRO,Aguilar2018OptionPM}. The fact that MB integrals occur in numerous different situations is certainly a strong motivation to develop and improve computational techniques that enable their analytic or numerical evaluation.

In the MB approach to compute Feynman integrals, the original scalar Feynman integral is first converted into an MB integral using automated tools such as \texttt{AMBRE}~\cite{Gluza:2007rt} and \texttt{MBcreate}~\cite{Belitsky:2022gba}, and then the MB integral is used for further analysis, such as resolving the $\epsilon$-singularities~\cite{Smirnov:1999gc,Tausk:1999vh,Czakon:2005rk,Smirnov:2009up}, evaluation in terms of hypergeometric series~\cite{Kalmykov:2020cqz} and special functions~\cite{Smirnov:2012gma,Vollinga:2004sn}, performing numerical integration~\cite{Czakon:2005rk,Dubovyk:2019krd}, counting master integrals in some cases~\cite{Kalmykov:2016lxx}, and deriving partial differential equations without relying on integration-by-parts identities~\cite{Kalmykov:2012rr}. 

In loop calculations, MB integrals have been used since the 1970s, when they were first used to compute one-loop three-point Feynman integrals in~\cite{Usyukina:1975yg}. They became popular after several works showing their efficiency in the 1990s \cite{Boos:1990rg, Davydychev:1990jt, Davydychev:1990cq,Usyukina:1992jd, Davydychev:1992mt}, and even more after an important breakthrough in the derivation of the first analytic results for two-loop box Feynman integrals in the planar~\cite{Smirnov:1999gc} and non-planar~\cite{Tausk:1999vh} cases.
More recently, MB integrals have been applied to the computation of the
single Higgs boson production cross-section at N$^3$LO~\cite{Anastasiou:2015vya,Anastasiou:2016cez,Anastasiou:2014lda}, the two-loop electroweak corrections to the $Z$-boson production and decay~\cite{Dubovyk:2018rlg}, NNLO corrections to Bhabha scattering~\cite{Banerjee:2021mty}, two-loop electro-weak corrections to Higgs boson pair production~\cite{Davies:2022ram,Davies:2023npk,Zhang:2024fcu,Davies:2025wke}, etc. This non-exhaustive list also includes low-energy precision physics works such as~\cite{Aguilar:2008qj,Greynat:2012ww,Charles:2017snx,Ananthanarayan:2017qmx}, studies of amplitudes in $\mathcal{N}=4$ SYM \cite{Bern:2005iz,Bern:2006ew}, computations of phase-space integrals~\cite{Somogyi:2011ir,Ahmed:2024pxr, Ahmed:2025yrx} and renormalization group equations for effective field theories~\cite{Banik:2025wpi}, among others.

Currently, the appearance of higher fold MB integrals for multi-scale multi-loop Feynman integrals limits the effectiveness of the MB approach for directly computing state-of-the-art Feynman integrals~\cite{Weinzierl:2022eaz}. Nevertheless, MB integrals remain very useful for computing Feynman integrals that are independent of kinematical variables (where the differential equation approach~\cite{Kotikov:1990kg,Gehrmann:1999as,Henn:2013pwa} cannot be used without introducing artificial scales), and for computing individual regions obtained from the method of regions~\cite{Beneke:1997zp,Pak:2010pt,Ananthanarayan:2018tog,Smirnov:2021dkb}. For multi-scale Feynman integrals, MB integrals remain useful, particularly for computing {boundary conditions} in the differential equation approach. Moreover, if, for a given Feynman integral, the MB representation has a dimension lower than the Feynman parameterization, it is more suitable for numerical integration~\cite{Usovitsch:2018shx}. Lastly, MB integrals are presently used to study formal properties of quantum field theory, such as in the context of conformal Feynman integrals \cite{,Ananthanarayan:2020xpd,Alkalaev:2025fgn, Alkalaev:2025zhg}, the link of the latter with Yangian symmetries \cite{Loebbert:2019vcj,Ananthanarayan:2020ncn},
correlators studies in de Sitter space
\cite{Choudhury:2025pzp},
heat-kernel expansion \cite{Barvinsky:2025pjh}, etc.

In recent years, two novel geometric techniques based on conic hulls~\cite{Ananthanarayan:2020fhl} and triangulations~\cite{Banik:2023rrz} have been developed to systematize the evaluation of MB integrals in terms of (multivariate) series, and the public software \texttt{MBConicHulls.wl}~\cite{MBConicHullsGit} has been developed. However, these techniques and, therefore, the package, are limited to MB integrals containing only gamma functions. In principle, this is sufficient if the singularity resolution for dimensionally regularized Feynman integrals is performed after evaluation, since the MB representation of unresolved convergent or dimensionally regularized Feynman integrals does not involve any polygamma functions. However, if one performs the $\epsilon$-resolution of MB integrals using standard methods~\cite{Czakon:2005rk,Smirnov:2009up} before evaluation, then polygamma functions will appear. This happens because expanding the MB integrand in $\epsilon$ requires taking derivatives with respect to $\epsilon$, and when these act on Euler gamma functions, they produce polygamma functions.

In this paper, we extend the conic hull and triangulation methods to handle MB integrals involving polygamma functions and to express them in terms of infinite series. These series can then be used for numerical studies and can sometimes also be rewritten in terms of special functions using symbolic summation tools~\cite{schneider2013}. To make our work ready-to-use, we also implement the new algorithms in an upgraded version of the package \texttt{MBConicHulls.wl} and provide some illustrative examples.

 The remainder of the article is structured as follows. In Section~\ref{MB_def}, we define our conventions and discuss how to compute MB integrals with polygamma functions, both using triangulation and conic hull approaches.
 In Section~\ref{MBstraight}, we discuss some complications in evaluating MB integrals with straight contours, and then show how one can handle such cases by rewriting the polygamma functions as derivatives of gamma functions. Next, in Section~\ref{MBConicHulls.wl}, we describe how, using \texttt{MBConicHulls.wl}, one can now compute MB integrals with polygamma functions by solving, as an example, one MB integral that appears in the Higgs production cross-section calculation. Finally, we present our concluding remarks in Section~\ref{Conclusion}. In Appendix~\ref{MBConicHull_Documentation}, we provide documentation of three external modules of~\texttt{MBConicHulls.wl} that have been updated since the last version.

\section{$N$-fold MB integrals with polygamma functions  \label{MB_def}}

In their usual form, $N$-fold MB integrals have the following general structure:
\begin{equation} 
    I(x_1,\cdots ,x_N)\equiv \int\limits_{-i \infty}^{+i \infty} \frac{ d z_1}{2 \pi i} \cdots \int\limits_{-i \infty}^{+i \infty}\frac{ d z_N}{2 \pi i}\,\,  \frac{\prod\limits_{i=1}^{k} \Gamma^{a_i}(s_i ({\bf z}))}{\prod\limits_{j=1}^{l} \Gamma^{b_j}(t_j ({\bf z}))}  x^{z_1}_{1} \cdots x^{z_N}_{N}\label{N_MB}
\end{equation}
where ${\bf z}=(z_1,\cdots,z_N)$, $a_i$ and $b_j$ are positive integers, $k\geq N${ (we tacitly exclude cancellations between numerator and denominator gamma functions)}  and the variables $x_1,\cdots ,x_N$ can be complex-valued. The $s_i$ and $t_j$ functions in the MB integrand are
\begin{align}\label{argument}
    s_i({\bf z}) =\sum\limits_{k=1}^{N}e_{i k}z_k+f_i \, , \hspace{2cm}
    t_j({\bf z}) =\sum\limits_{k=1}^{N}g_{j k}z_k+h_j
\end{align}
where $f_i$ and $h_j$ are real or complex numbers, and the coefficients $e_{ik}$ and $g_{jk}$ are often integers. 

If not otherwise stated, the contours of integration in Eq. (\ref{N_MB}) do not split the set of poles of each gamma function of the numerator in different subsets. 
As indicated in \cite{Banik:2022bmk} it is sometimes convenient to rewrite Eq. \eqref{N_MB} in the canonical form
\begin{equation} \label{N_MB_2}
    I(x_1,\cdots ,x_N)= \int\limits_{-i \infty}^{+i \infty} \frac{ d z_1}{2 \pi i} \cdots \int\limits_{-i \infty}^{+i \infty}\frac{ d z_N}{2 \pi i}\,\,  \frac{ \Gamma(-z_1)\cdots\Gamma(-z_N) \prod\limits_{i=N+1}^{k'} \Gamma^{a'_i}(s'_i ({\bf z})) }{\prod\limits_{j=1}^{l} \Gamma^{b'_j}(t'_j ({\bf z}))} x'^{z_1}_{1} \cdots x'^{z_N}_{N}
\end{equation}
where 
\begin{align}\label{argument_canonical}
    s'_i({\bf z}) =\sum\limits_{k=1}^{N}e'_{i k}z_k+f'_i \, , \hspace{2cm}
    t'_j({\bf z}) =\sum\limits_{k=1}^{N}g'_{j k}z_k+h'_j
\end{align}
In this paper, however, we are interested in the more general situation where polygamma functions $\psi(m,z)$, with non-negative integer $m$, also appear in the numerator of the integrand\footnote{We do not consider polygamma functions in the denominator, since such MB integrals do not arise in Feynman integral calculations.}. Therefore, our goal is to evaluate MB integrals of the following form
\begin{align} \label{N_MB_3}
   & J(x_1,\cdots ,x_N)=\nonumber \\
    &\int\limits_{-i \infty}^{+i \infty} \frac{ d z_1}{2 \pi i} \cdots \int\limits_{-i \infty}^{+i \infty}\frac{ d z_N}{2 \pi i}\,\,  \frac{ \Gamma(-z_1)\cdots\Gamma(-z_N) \prod\limits_{i=1}^{m} \Gamma^{a_i}(s_i ({\bf z}))\prod\limits_{p=1}^{n} \psi^{c_p}(m_p,u_p ({\bf z})) }{\prod\limits_{j=1}^{l} \Gamma^{b_j}(t_j ({\bf z}))} x^{z_1}_{1} \cdots x^{z_N}_{N}
\end{align}
where  $ u_p ({\bf z})=\sum\limits_{k=1}^{N}q_{p k}z_k+r_p $ and, as in the case of  $s_i({\bf z})$ and $t_j({\bf z})$, $r_p$ are real or complex numbers, while coefficients $q_{pk}$  are often integers.

By definition, the polygamma function $\psi(m,z)$ is given by the $(m+1)$-th derivative of the logarithm of the gamma function
\begin{equation}
    \psi(m,z) = \frac{d^{\,m+1}}{dz^{\,m+1}} \ln \Gamma(z)\,.
\end{equation}
Hence, like the gamma function $\Gamma(z)$, the polygamma functions $\psi(m,z)$ have poles at all non-positive integers. However, the multiplicity of these poles depends on the order $m$ of the polygamma function.

MB integrals of the form in Eq.~\eqref{N_MB_3} can be systematically evaluated using both conic-hull and triangulation techniques. To do this, one should treat the argument of the polygamma function in the same way as the gamma function arguments in the numerator to construct conic hulls or point configurations in the conic-hull and triangulation methods, respectively. In this way, different sets of poles (also known as cones~\cite{Friot:2011ic}), whose sums of residues yield different series solutions, which are analytic continuations of each other, are identified.

However, in the next step, the evaluation of multivariate residues is different from before 
because the singularity structure of the polygamma function differs from that
of the gamma function. To extract the residues at the poles of the polygamma
functions, we first apply the generalized reflection formula
\begin{align} 
\psi(m,z-n)=\frac{(-1)^{m+1}m!}{z^{m+1}}+\psi(m,1+z)+(-1)^{m+1}\psi(m,1-z) + (-1)^m \psi(m,1+n-z)\,,
\label{psi}
\end{align}
which can be derived from the well-known reflection formula for the gamma function
\begin{align} 
\Gamma(z-n)=\frac{\Gamma(z+1)\Gamma(1-z)(-1)^n}{z\ \Gamma(n+1-z)}\,,
\label{reflection}
\end{align}
valid for any non-negative integer $n$. We then group the singular factors in a similar way as discussed in~\cite{Ananthanarayan:2020fhl}, and finally compute the residues using the package \texttt{MultivariateResidues.m} ~\cite{Larsen:2017aqb}, which is internally called by \texttt{MBConicHulls.wl}.

\section{Limiting approach for MB integrals with straight contours \label{MBstraight}}

Although the conic hull and triangulation methods work for any MB integral with polygamma functions, there are complications in evaluating MB integrals with straight contour splitting the sets of poles of some polygamma functions.

To understand why this is the case, let us consider the following toy two-fold MB integral
\begin{align} \label{MB_straight}
    I &(x_1,x_2) = \int\limits_{c_1-i \infty}^{c_1+i \infty} \frac{ \text{d} z_1}{2 \pi i} \int\limits_{c_2-i \infty}^{c_2+i \infty}\frac{ \text{d} z_2}{2 \pi i}\,\, (-x_1)^{z_1} (-x_2)^{z_2} \Gamma(-z_1)\Gamma(-z_2) 
    \Gamma \left(1+z_1+z_2 \right) \psi \left(1, 1+z_1+z_2 \right)
\end{align}
where the straight contours are chosen such that $c_1 \doteq \Re(z_1)= -\frac{7}{9}$ and $c_2 \doteq \Re(z_2)= -\frac{3}{5}$. For this integral, we note that the sets of poles of
$\Gamma\!\left(1+z_{1}+z_{2}\right)$ and
$\psi\!\left(1,1+z_{1}+z_{2}\right)$ are split by the integration
contours, because the contours are chosen such that the real part of the
argument of each of these two functions is negative along them. In such a scenario, the conic hull and triangulation methods do not work, as noted in~\cite{Banik:2022bmk}, unless one rewrites the MB integrand in such a way that the contours do not split the sets of poles of any of the numerator gamma or polygamma functions anymore.

For the gamma function $\Gamma \left(1+z_1+z_2 \right)$, this is easy to do by rewriting it as 
\begin{align}\label{straight2nonstraightGamma}
\Gamma\left(1+z_1+z_2\right)
  = -\frac{\Gamma(2+z_1+z_2)\Gamma(-1-z_1-z_2)}{\Gamma(-z_1-z_2)}\,,
\end{align}
following the strategy presented in~\cite{Banik:2022bmk}, such that now all new numerator gamma-function arguments have positive real parts along the contour. However, for the polygamma function $\psi \left(1, 1+z_1+z_2 \right)$, this is possible only if it is written as a sum of three terms,
\begin{align}\label{straight2nonstraightPolyGamma}
\psi \left(1, 1+z_1+z_2 \right)
  = - \psi \left(1, 2+z_1+z_2 \right)
    - \psi \left(1, -1-z_1-z_2 \right)
    + \psi \left(1, -z_1-z_2 \right)\,,
\end{align}
and therefore we need to evaluate three MB integrals separately with different conic-hull and triangulation structures,
\begin{align} \label{MB_straight_2}
    I(x_1,x_2) &=
    \int\limits_{c_1-i \infty}^{c_1+i \infty} \frac{ \mathrm{d} z_1}{2 \pi i}
    \int\limits_{c_2-i \infty}^{c_2+i \infty}\frac{ \mathrm{d} z_2}{2 \pi i}
    \, (-x_1)^{z_1} (-x_2)^{z_2} 
    \frac{\Gamma(-z_1)\Gamma(-z_2) \Gamma(2+z_1+z_2)\Gamma(-1-z_1-z_2)}{\Gamma(-z_1-z_2)}
\nonumber \\[0.3em]
 &\hspace{2em}\times
 \Bigl[
   - \psi \left(1, 2+z_1+z_2 \right)
   - \psi \left(1, -1-z_1-z_2 \right)
   + \psi \left(1, -z_1-z_2 \right)
 \Bigr] \,.
\end{align}

Thus, after solving each of the three MB integrals, we need to relate their different series solutions and sum the appropriate ones to obtain the correct final result for $I(x_1,x_2)$. It turns out that each of the three MB integrals in Eq.~\eqref{MB_straight_2} has four different series solutions, so in total we obtain 12 series solutions. Now, to obtain the series solutions of $I(x_1,x_2)$, these 12 series must be grouped into various triplets, where each triplet corresponds to a series solution for $I(x_1,x_2)$. 

Determining these groupings, however, requires knowing in advance the convergence region of each series solution of the three individual MB integrals, which, if still doable for the simple case at hand, is highly non-trivial for higher-fold hypergeometric series.
Therefore, to bypass this complication, whenever we have polygamma functions whose poles are split by the contours, we first rewrite them as
\begin{equation}\label{PG2G}
    \psi(0,z) = \lim_{a \to 0}\frac{d}{da} \left[ \frac{\Gamma(z+a)}{\Gamma(z)} \right], 
    \hspace{1cm} 
    \psi(m,z) = \lim_{\substack{a \to 0 \\[0.1cm] b \to 0}} \, \frac{d^{\,m}}{db^{\,m}} \frac{d}{da} \left[ \frac{\Gamma(z+a+b)}{\Gamma(z+b)} \right]
    \quad \text{for } m \geq 1 
\end{equation}
by introducing one or two auxiliary variables $a$ and $b$, depending on the order of the polygamma function. One then has only gamma functions in the integrand, which for Eq.~\eqref{MB_straight} reads
\begin{align} \label{MB_straight_3}
    I^\prime (x_1,x_2,a,b) = \int\limits_{c_1-i \infty}^{c_1+i \infty} \frac{ \text{d} z_1}{2 \pi i} \int\limits_{c_2-i \infty}^{c_2+i \infty}\frac{ \text{d} z_2}{2 \pi i}\,\, x_1^{z_1} x_2^{z_2} \,\, &\Gamma(-z_1)\Gamma(-z_2) 
    \Gamma \left(1+z_1+z_2 \right)  \nonumber \\& \times
    \frac{\Gamma \left(1+a+b+z_1+z_2 \right)}{\Gamma \left(1+b+z_1+z_2 \right)}
\end{align}
from which  $I (x_1,x_2)$ can be derived using 
\begin{align} \label{MB_limit}
    I &(x_1,x_2) = \lim_{\substack{a \to 0 \\[0.1cm] b \to 0}} \, \frac{d}{db} \frac{d }{da} I^\prime(x_1,x_2,a,b)
\end{align}
and therefore, in the next step, we can rewrite gamma functions whose poles are split by the contour in a product form as in Eq.~\eqref{straight2nonstraightGamma} such that the contours do not split their poles anymore.

\begin{align} \label{MB_straight_4}
    I^\prime (x_1,x_2,a,b) = \int\limits_{c_1-i \infty}^{c_1+i \infty} \frac{ \text{d} z_1}{2 \pi i} \int\limits_{c_2-i \infty}^{c_2+i \infty}\frac{ \text{d} z_2}{2 \pi i}&\,\, x_1^{z_1} x_2^{z_2} \,\, \frac{\Gamma(-z_1)\Gamma(-z_2) \Gamma(2+z_1+z_2)\Gamma(-1-z_1-z_2)}{\Gamma(-z_1-z_2)} \nonumber \\  & \times
    \frac{\Gamma \left(2+a+b+z_1+z_2 \right) \Gamma \left(-1-a-b-z_1-z_2 \right)}{\Gamma \left(-a-b-z_1-z_2 \right) \Gamma \left(1+b+z_1+z_2 \right)}
\end{align}

The resulting integrand is then in a suitable form for applying the conic-hull or triangulation methods in order to identify the different sets of poles that contribute to the various series solutions. We note here that the introduction of the parameters $a$ and $b$ does not present problems, as it does not modify the pole structure of the integrand. Indeed, we recall that the conic-hull and triangulation methods, in the non-straight contour case, can give results even for non fixed values of the constants $f_i$, $h_j$ and $r_p$ (see Section \ref{MB_def} for their definition) as long as the latter are considered generic, which is the case in our way to rewrite polygamma functions in terms of derivatives of gamma functions. 

After this step, we carry out the derivatives and limits in Eq.~\eqref{MB_limit} {and then compute the residues} to obtain the final result for $I(x_1,x_2)$. More details on this example are given in the accompanying notebook
\texttt{Examples\_PolyGamma.nb} in~\cite{MBConicHullsGit}, where we compute Eq.~\eqref{MB_straight} in two
different ways: first by rewriting the polygamma function according to
Eq.~\eqref{straight2nonstraightPolyGamma}, and second by expressing it in terms
of gamma functions using Eq.~\eqref{PG2G}. We showed in the notebook that both approaches lead to
the same final result, although in practice the representation in Eq.~\eqref{PG2G} is
preferable as discussed above.

We note that the procedure of rewriting polygamma functions in terms of gamma functions can also be applied in the non-straight contour case. However, we found that, in practice, evaluating the derivatives and limits is computationally more expensive than computing with the polygamma function directly. For this reason, in the \texttt{MBConicHulls.wl} package, as discussed next, we use Eq.~\eqref{PG2G} only for polygamma functions whose poles are split by the straight contours.

\section{Implementation in \texttt{MBConicHulls.wl} and illustrative examples  \label{MBConicHulls.wl}}

In this section, we discuss new developments in \texttt{MBConicHulls.wl} that enable the evaluation of MB integrals with polygamma functions. \texttt{MBConicHulls.wl} is a \textit{Mathematica} package that automates the analytic evaluation of $N$-fold MB integrals in terms of multivariate series. It was originally built on the conic-hull method and was later extended to incorporate the triangulation method as well. The latest version of this package, along with tutorial notebooks featuring pedagogical examples, can be downloaded from the following \texttt{GitHub} repository.

\begin{center}
    \href{https://github.com/SumitBanikGit/MBConicHulls}{\texttt{https://github.com/SumitBanikGit/MBConicHulls}}
\end{center}

In the following, we explicitly compute a two-fold MB integral to illustrate the usage and new features of \texttt{MBConicHulls.wl}. We also discuss additional examples that we have computed to validate our computer implementation.

\subsection{Master integral appearing in single-Higgs production at N$^3$LO \label{MI_N3LO}}

In order to show how one can use \texttt{MBConicHulls.wl} to evaluate MB integrals with polygamma functions, we take the following two-fold MB integral~\cite{Anastasiou:2013srw}

\begin{align}\label{F82}
	\mathcal{F}_{8,2} &=\int\limits_{c_3-i \infty}^{c_3+i \infty}  \frac{dz_3}{2 \pi i} \int\limits_{c_4-i \infty}^{c_4+i \infty}  \frac{dz_4}{2 \pi i} \frac{\Gamma(-z_3)^3\Gamma(-z_4)^3\Gamma(z_3)\Gamma(z_4)\Gamma(1+z_3)\Gamma(1+z_4)}{2 \, \epsilon \, \Gamma(-z_3-z_4)} \nonumber \\[0.3cm] &
	\times[3\epsilon\,\psi(0,-z_3)+\epsilon\,\psi(0,z_3)-2\epsilon\,\psi(0,-z_3-z_4)+\epsilon\,\psi(0,-z_4)+\epsilon\,\psi(0,z_4)-1]
\end{align}

with $c_3 = -0.64 $ and $c_4 = -0.22 $. The above integral $\mathcal{F}_{8,2}$ is obtained after performing the $\epsilon$-resolution of the master integral $\mathcal{F}_{8}$, which is one of the master integrals that contributes to the soft triple-radiation part of single-Higgs production cross-section at N$^3$LO (see~\cite{Anastasiou:2013srw} for more details).

Since the MB integral in Eq.~\eqref{F82} is symmetric under the exchange of $z_3$ and $z_4$ (up to the 3 factor in front of the first polygamma function), we need to evaluate only the following four integrals

\begin{align}\label{F821}
	\mathcal{F}_{8,2,1} &=\int_{-i \infty}^{+i \infty}  \frac{dz_3}{2 \pi i}\int_{-i \infty}^{+i \infty}  \frac{dz_4}{2 \pi i} \frac{\Gamma(-z_3)^3\Gamma(-z_4)^3\Gamma(z_3)\Gamma(z_4)\Gamma(1+z_3)\Gamma(1+z_4)}{ \Gamma(-z_3-z_4)} 
\end{align}
\begin{align}\label{F822}
	\mathcal{F}_{8,2,2} &=\int_{-i \infty}^{+i \infty}  \frac{dz_3}{2 \pi i}\int_{-i \infty}^{+i \infty}  \frac{dz_4}{2 \pi i} \frac{\Gamma(-z_3)^3\Gamma(-z_4)^3\Gamma(z_3)\Gamma(z_4)\Gamma(1+z_3)\Gamma(1+z_4)}{\Gamma(-z_3-z_4)} \psi(0,-z_3)
\end{align}
\begin{align}\label{F823}
	\mathcal{F}_{8,2,3} &=\int_{-i \infty}^{+i \infty}  \frac{dz_3}{2 \pi i}\int_{-i \infty}^{+i \infty}  \frac{dz_4}{2 \pi i} \frac{\Gamma(-z_3)^3\Gamma(-z_4)^3\Gamma(z_3)\Gamma(z_4)\Gamma(1+z_3)\Gamma(1+z_4)}{\Gamma(-z_3-z_4)} \psi(0,z_3)
\end{align}
\begin{align}\label{F824}
	\mathcal{F}_{8,2,4} &=\int_{-i \infty}^{+i \infty}  \frac{dz_3}{2 \pi i}\int_{-i \infty}^{+i \infty}  \frac{dz_4}{2 \pi i} \frac{\Gamma(-z_3)^3\Gamma(-z_4)^3\Gamma(z_3)\Gamma(z_4)\Gamma(1+z_3)\Gamma(1+z_4)}{\Gamma(-z_3-z_4)} \psi(0,-z_3-z_4)
\end{align}
to compute $\mathcal{F}_{8,2}$ using
\begin{equation}\label{F82_dec}
    \mathcal{F}_{8,2} = -\frac{\mathcal{F}_{8,2,1}}{2 \epsilon} + 2\,\mathcal{F}_{8,2,2} + \mathcal{F}_{8,2,3} - \mathcal{F}_{8,2,4}
\end{equation}
We now show, step by step, how to compute the integral $\mathcal{F}_{8,2,3}$ using \texttt{MBConicHulls.wl}. The remaining integrals are computed in the accompanying \texttt{Examples\_PolyGamma.nb} notebook in~\cite{MBConicHullsGit}.

First, we load the \texttt{MBConicHulls.wl} package
\mybox{
\inbox{1}{& {\tt Get["MBConicHulls.wl"];}}\\
\boxsplit\\\\
\outbox{2}{& {\tt Last Updated: 12th December, 2025
}\\
& {{\tt Version 1.3.4 by S.Banik \& S. Friot}}}
}
and then enter the MB in Eq.~\eqref{F823} using the \texttt{MBRep[]} module as shown below.
\mybox{
\inbox{2}{& ${\tt F823Rep = MBRep\Big[1,\ \{z_3 \to -0.64,\ z_4 \to -0.22\},\ \{u_1,u_2\},}$}\\
\gibspace{}{& ${\tt \qquad
\Big\{\{-z_3,-z_3,-z_3,-z_4,-z_4,-z_4,z_3,z_4,1+z_3,1+z_4,\{0,z_3\}\},} \{\tt-z_3-z_4\} \Big\}\Big]; $}\\
\boxsplit\\
\outbox{1}{& ${\tt  Straight\ Contour:\ \{\Re[z_3] = -0.64,\ \Re[z_4] = -0.22\}}$\\
& ${\tt Transforming\ PolyGamma\ functions\ whose\ poles\ are\ split\ by\ the\ straight\ }$\\
& ${\tt contours\ into\ Gamma\ functions}$\\
& ${\tt Introduced\ Dummy\ variable:\ \{\iota_{11}\}}$\\
& ${\tt Time\ Taken\ 0.026536\ seconds}$\\
}\\
}

Although our integral in Eq.~\eqref{F823} has no scales, we introduced the two artificial scales $u_1$ and $u_2$ so that \texttt{ResolveMB[]} can find the master series. However, if finding the master series is not relevant to the user, the artificial scales do not need to be introduced.
Nevertheless, these scales will be absent from the final result, as we will set them to one prior to residue evaluation, as discussed later in this section.

 In passing by, we note here two new features of \texttt{MBRep[]}. First, to input a polygamma function of the form $\psi(m,z)$, one should use a list with two elements, the first being the order and the second being the argument of the polygamma function. Thus, in the above case, we used $\texttt{\{0,$\tt z_3$\}}$ to input $\psi(0,z_3)$ in the numerator. The second new feature is the introduction of auxiliary variables, which are necessary to rewrite polygamma functions (only those whose poles are split by the straight contour) in terms of gamma functions using Eq.~\eqref{PG2G}. We give the user the choice not to transform polygamma functions to gamma functions, but rather to use a linear equation similar to Eq.~\eqref{straight2nonstraightPolyGamma} by setting $\texttt{PolygammaToGamma $\to$ False}$. However, this can lead to complications as discussed in the previous section, so in this section we keep $\texttt{PolygammaToGamma $\to$ True}$; which is also the default option of \texttt{MBRep[]}.

Our next step is to identify the different sets of poles that contribute to the various series solutions. For this, one can use either \texttt{TriangulateMB[]} to apply the triangulation method or \texttt{ResolveMB[]} to apply the conic-hull method. Both modules always produce the same set of series solutions (but not extracted in the same order), although the triangulation method is often much faster for higher-dimensional MB integrals, as noted in~\cite{Banik:2023rrz}. Since we are dealing with a two-fold MB integral, this choice is not
crucial, and we therefore use the conic-hull method by calling
\texttt{ResolveMB[]} with \texttt{PrintSolutions $\to$ False} to keep the
output minimal.
\vspace{-0.5cm}
\mybox{
\inbox{3}{& ${\tt F823Resolve = ResolveMB\Big[F823Rep,\, PrintSolutions \to False\Big]; }$}\\
\boxsplit\\
\outbox{1}{& ${\tt Degenerate \,\, case \,\, with \,\, 36 \,\, conic \,\, hulls }$\\
& ${\tt Found \,\, 4 \,\, series \,\, solutions. }$\\
& ${\tt Cardinality \,\, 4 :: \,\, Solution \,\, found \,\, 1. }$\\
& ${\tt Cardinality \,\, 8 :: \,\, Solution \,\, found \,\, 2. }$\\
& ${\tt Cardinality \,\, 16 :: \,\, Solution \,\, found \,\, 1. }$\\[8pt]
& ${\tt Time \,\, Taken \,\, 0.203493 \,\, seconds }$\\
}\\
}

We observe that $\mathcal{F}_{8,2,3}$ has 36 associated conic hulls and that four distinct series solutions were found. Among these, the first solution is the shortest with cardinality 4, and so we next derive the explicit analytic expression of this solution.
\vspace{-0.5cm}
\mybox{
\inbox{4}{
& ${\tt SeriesNumber = 1;}$\\
& ${\tt F823Series = EvaluateSeries\Big[ F823Resolve,\ \{u_1 \to 1,\ u_2 \to 1\},\ SeriesNumber \Big]; }$
}\\
\boxsplit\\
\outbox{1}{
& ${\tt The \,\, series \,\, solution \,\, is \,\, a \,\, sum \,\, of \,\, the \,\, following \,\, 1 \,\, series. }$\\[8pt]

& \hspace{-1.4cm} ${\tt \textbf{Series Number 1} ::}$\\[4pt]
& $\scriptsize
\begin{aligned}[t]
 &\hspace{-1.4cm} (-1)^{2(n_{1}+n_{2})}\,\Gamma(1+n_{1})^{2}\,\Gamma(1+n_{2})^{2} 
 \times\Bigl(
 2\,\mathrm{PolyGamma}(0,1+n_{2})\,\mathrm{PolyGamma}(0,2+n_{1}+n_{2})^{2}\\[2pt]
 & \hspace{-1.4cm}
 -\mathrm{PolyGamma}(0,2+n_{2})\,\mathrm{PolyGamma}(0,2+n_{1}+n_{2})^{2}
 -\mathrm{PolyGamma}(0,2+n_{1}+n_{2})^{3} \\
 &\hspace{-1.4cm} 
 +\mathrm{PolyGamma}(0,2+n_{1})^{2}
   \bigl(
      -2\,\mathrm{PolyGamma}(0,1+n_{2})
      +\mathrm{PolyGamma}(0,2+n_{2})
      +\mathrm{PolyGamma}(0,2+n_{1}+n_{2})
   \bigr)\\
 &\hspace{-1.4cm}
 +\mathrm{PolyGamma}(0,1+n_{1})^{2}
   \bigl(
      8\,\mathrm{PolyGamma}(0,1+n_{2})
      -4\bigl(
          \mathrm{PolyGamma}(0,2+n_{2})
         +\mathrm{PolyGamma}(0,2+n_{1}+n_{2})
        \bigr)
   \bigr)\\
 & \hspace{-1.4cm}
 +4\,\mathrm{PolyGamma}(0,1+n_{2})\,\mathrm{PolyGamma}(1,1+n_{1})
 -2\,\mathrm{PolyGamma}(0,2+n_{2})\,\mathrm{PolyGamma}(1,1+n_{1}) \\
 &\hspace{-1.4cm}
 -2\,\mathrm{PolyGamma}(0,2+n_{1}+n_{2})\,\mathrm{PolyGamma}(1,1+n_{1})
 +2\,\mathrm{PolyGamma}(0,1+n_{2})\,\mathrm{PolyGamma}(1,2+n_{1}) 
 \\
 & \hspace{-1.4cm}
 -\mathrm{PolyGamma}(0,2+n_{2})\,\mathrm{PolyGamma}(1,2+n_{1})
 -\mathrm{PolyGamma}(0,2+n_{1}+n_{2})\,\mathrm{PolyGamma}(1,2+n_{1}) \\
 &\hspace{-1.4cm}
 -2\,\mathrm{PolyGamma}(0,1+n_{2})\,\mathrm{PolyGamma}(1,2+n_{1}+n_{2})
 +\mathrm{PolyGamma}(0,2+n_{2})\,\mathrm{PolyGamma}(1,2+n_{1}+n_{2}) \\
 &\hspace{-1.4cm}
 +3\,\mathrm{PolyGamma}(0,2+n_{1}+n_{2})\,\mathrm{PolyGamma}(1,2+n_{1}+n_{2}) \\
 &\hspace{-1.4cm}
 -4\,\mathrm{PolyGamma}(0,1+n_{1})\bigl(
     2\,\mathrm{PolyGamma}(0,1+n_{2})\,\mathrm{PolyGamma}(0,2+n_{1}+n_{2})
     \\
 &\hspace{-1.4cm}
 -\mathrm{PolyGamma}(0,2+n_{2})\,\mathrm{PolyGamma}(0,2+n_{1}+n_{2})
    -\mathrm{PolyGamma}(0,2+n_{1}+n_{2})^{2}
    +\mathrm{PolyGamma}(1,2+n_{1}+n_{2})
   \bigr)\\
 & \hspace{-1.4cm}
 -\mathrm{PolyGamma}(2,2+n_{1}+n_{2})
 \Bigr) / \Bigl( 2\,\Gamma(2+n_{1})\,\Gamma(2+n_{2})\,\Gamma(2+n_{1}+n_{2}) \Bigr) 
 \\
 & 
\end{aligned}
$  \\[4pt]
& \hspace{-1.4cm} {\tt valid  for } $\,n_{1} \ge 0 \,\,\&\&\,\, n_{2} \ge 0$\\[8pt]

& \hspace{-1.4cm} ${\tt Time \,\, Taken \,\, 0.695257 \,\, seconds }$\\
}
}

Note that in the above, we have set the artificial scales $u_1$ and $u_2$ to one before deriving the analytic solution. {The final result consists of a single two-fold infinite series, although the cardinality of the solution was four. This is because cancellation of poles from the denominator gamma functions caused the residues associated with three of the four sets of poles to vanish.} The remaining series is somewhat cumbersome, as seen above. Fortunately, we can rewrite it in terms of multiple zeta values using the \texttt{EvaluateMultiSum}~\cite{schneider2013} package

\mybox{
\inbox{5}{& ${\tt F823Val = EvaluateMultiSum[F823Series[[2, 1]], \{\{n_1, 0, \infty\}, \{n_2, 0, \infty\}\}] // Expand}$}\\
\boxsplit\\
\out{1}{
& ${\tt - \dfrac{6}{5}\, z2^{3} \,+\, 2\,EulerGamma\, z2\, z3 \,+\, 2\, z3^{2} \,-\, 9\,EulerGamma\, z5}$\\
}
}
where, in the above output, $z2 , z3, z5$ is $\zeta(2), \zeta(3), \zeta(5) $, respectively. We follow the exact same procedure above to compute the remaining MB integrals in Eqs.~\eqref{F821},~\eqref{F822},~\eqref{F824} using \texttt{MBConicHulls.wl} to obtain

\begin{align}
    \mathcal{F}_{8,2,1} & = -2 \, \zeta_2 \zeta_3 + 9 \, \zeta_3 \, ,\\
    \mathcal{F}_{8,2,2} & =  -\frac{41}{105}\,\zeta_{2}^{3}
+ 2\,\gamma_{\mathrm{E}}\,\zeta_{2}\,\zeta_{3}
- \zeta_{3}^{2}
- 9\,\gamma_{\mathrm{E}}\,\zeta_{5}\,,
\\
    \mathcal{F}_{8,2,4} & = \frac{401}{210}\,\zeta_{2}^{3}
+ 2\,\gamma_{\mathrm{E}}\,\zeta_{2}\,\zeta_{3}
- 4\,\zeta_{3}^{2}
- 9\,\gamma_{\mathrm{E}}\,\zeta_{5}
\end{align}
and using Eq.~\eqref{F82_dec}, we obtain the final result
\begin{align}
    \mathcal{F}_{8,2}
    &= -\frac{817}{45360}\,\pi^{6}
    + \frac{2}{3}\,\gamma_{\mathrm{E}}\,\pi^{2}\,\zeta_{3}
    + 4\,\zeta_{3}^{2}
    - 18\,\gamma_{\mathrm{E}}\,\zeta_{5}
    + \frac{1}{\epsilon}\!\left(\frac{\pi^{2}}{6}\,\zeta_{3}
    - \frac{9}{2}\,\zeta_{5}\right)\,,
\end{align}
which agrees with Eq.~(8.61) in Ref.~\cite{Anastasiou:2013srw}. Therefore, this concludes the example; further details are provided in the accompanying \textit{Mathematica} notebook \texttt{Examples\_PolyGamma.nb}.

In addition to the example discussed above, we have validated our package with other integrals as well. One of them is the two-fold MB representation of the double-box integral~\footnote{We have omitted the constant factor $\frac{1}{x_{12}^{2}}$ here since it does not affect our subsequent analysis.} in Eq.~(50) of~\cite{Eden:2025frf}
\begin{align}\label{db_MB}
\mathcal{I}_{2} ( u , v )
  &= \frac{1}{2}
\int_{-i \infty}^{+i \infty}  \frac{dz_4}{2 \pi i}\int_{-i \infty}^{+i \infty}  \frac{dz_6}{2 \pi i} \,
        u^{z_{4}} v^{z_{6}} \,
        \Gamma(-z_{4})^{2}\Gamma(-z_{6})^{2}\Gamma(1+z_{4}+z_{6})^{2}
\nonumber \\[0.5em]
  &\qquad \times
     \Bigl(\pi^{2}
           + \bigl(\psi(0,-z_{4})-\psi(0,-z_{6})\bigr)^{2}
           - \psi(1,-z_{4}) - \psi(1,-z_{6})\Bigr)\,
\end{align}
which has non-straight contours here but could be written with straight contours as well (indeed it is possible to find contours such that $\Re (-z_4)>0$, $\Re (-z_6)>0$ and $\Re(1+z_4+z_6)>0$ simultaneously). Since the above integral is symmetric under $z_4 \leftrightarrow z_6$ and $u\leftrightarrow v$, we need to compute only four distinct integrals after expanding the terms in parentheses in Eq.~\eqref{db_MB}:
\begin{align}\label{I21}
	\mathcal{I}_{2,1} (u, v) &=\int_{-i \infty}^{+i \infty}  \frac{dz_4}{2 \pi i}\int_{-i \infty}^{+i \infty}  \frac{dz_6}{2 \pi i} \,
        u^{z_{4}} v^{z_{6}} \,
        \Gamma(-z_{4})^{2}\Gamma(-z_{6})^{2}\Gamma(1+z_{4}+z_{6})^{2} \,,
\end{align}
\begin{align}\label{I22}
	\mathcal{I}_{2,2} (u, v) &=\int_{-i \infty}^{+i \infty}  \frac{dz_4}{2 \pi i}\int_{-i \infty}^{+i \infty}  \frac{dz_6}{2 \pi i} \,
        u^{z_{4}} v^{z_{6}} \,
        \Gamma(-z_{4})^{2}\Gamma(-z_{6})^{2}\Gamma(1+z_{4}+z_{6})^{2} \, \psi(0,-z_{4})^{2} \,,
\end{align}
\begin{align}\label{I23}
	\mathcal{I}_{2,3} (u, v) &=\int_{-i \infty}^{+i \infty}  \frac{dz_4}{2 \pi i}\int_{-i \infty}^{+i \infty}  \frac{dz_6}{2 \pi i} \,
        u^{z_{4}} v^{z_{6}} \,
        \Gamma(-z_{4})^{2}\Gamma(-z_{6})^{2}\Gamma(1+z_{4}+z_{6})^{2} \,
        \psi(0,-z_{4}) \, \psi(0,-z_{6}) \,,
\end{align}
\begin{align}\label{I24}
	\mathcal{I}_{2,4} (u, v) &=\int_{-i \infty}^{+i \infty}  \frac{dz_4}{2 \pi i}\int_{-i \infty}^{+i \infty}  \frac{dz_6}{2 \pi i} \,
        u^{z_{4}} v^{z_{6}} \,
        \Gamma(-z_{4})^{2}\Gamma(-z_{6})^{2}\Gamma(1+z_{4}+z_{6})^{2} \, \psi(1,-z_{4}) \,,
\end{align}

We used \texttt{MBConicHulls.wl} to
analytically evaluate the four MB integrals above and found three series solutions for each. We then grouped them according to their master series characteristic list and inserted them into the master formula
\begin{equation}\label{db_master}
    \mathcal{I}_{2} ( u , v ) = \frac{1}{2} \bigg[ \pi^2  \,\mathcal{I}_{2,1} (u,v)
    + \mathcal{I}_{2,2} (u,v) + \mathcal{I}_{2,2} (v,u)
    - 2 \, \mathcal{I}_{2,3} (u,v)
    - \mathcal{I}_{2,4} (u,v)
    - \mathcal{I}_{2,4} (v,u)\bigg]
\end{equation}
to compute the original integral $I_{2}(u,v)$. Our final result matched perfectly with the direct numerical integration using \texttt{MB.m}. Therefore, this provides another non-trivial check of our code; further details for this example are given in \texttt{Examples\_PolyGamma.nb} in~\cite{MBConicHullsGit}. 

In addition to the above two examples, we also
computed the specific configuration $\mathbf{J011}(1,2,2)$ of the two-loop
sunset diagram\footnote{For a recent study of sunset diagrams with the MB approach see~\cite{Ananthanarayan:2025rti}.} studied in Ref.~\cite{Davydychev:2000na} and found excellent agreement until order $\mathcal{O}(\epsilon^2)$ (analytically at order $\mathcal{O}(\epsilon)$)~\cite{Banik:2025inprep}.

\section{Conclusion \label{Conclusion}}
The MB computational technique is a powerful tool used in high-energy physics and other domains of theoretical physics and mathematics to evaluate complicated integrals. In this paper, we have updated two geometric approaches, based on conic hulls and triangulations, to compute MB integrals with polygamma functions for both non-straight and straight contours, and have discussed how some complications that arise in the latter case can be solved following a simple approach alternative to the one used in the non-straight case. 

To make our work ready-to-use for practical calculations, we have implemented our algorithm in an upgraded version of the \textit{Mathematica} package \texttt{MBConicHulls.wl} and have illustrated its usage by explicitly showing each step of the computation of a two-fold MB integral with polygamma functions, which appears in the single-Higgs production cross-section calculation for the LHC. 

Taken together, our results now enable the (in principle) fully automated analytic evaluation of MB integrals arising from dimensionally regularised Feynman integrals. A possible workflow is as follows: first, use the packages \texttt{AMBRE} or \texttt{MBcreate} to derive the MB representation of a Feynman integral; second, use the packages~\texttt{MB.m} or~\texttt{MBresolve.m} to perform the $\epsilon$-resolution; then evaluate the resulting MB integrals with~\texttt{MBConicHulls.wl}; and finally use\footnote{This step may not always succeed and is currently a promising direction for future research.} \texttt{EvaluateMultiSum} to obtain a closed-form result. In the future, we plan to interface all these tools into a single package for a fully automated evaluation of MB integrals. 

\section*{Acknowledgements}
SB acknowledges support from the University of Zurich Postdoc Grant (Grant No.~FK-24-100) and Swiss National Science Foundation (Grant No.~PP00P21 76884).

\appendix

\section{Updated modules of \texttt{MBConicHulls.wl} \label{MBConicHull_Documentation}}

To implement the evaluation of MB integrals with polygamma functions in \texttt{MBConicHulls.wl}, we have updated the following three external modules whose documentation is provided below. All other external modules that are not described below remain unchanged.

\subsection{\texttt{MBRep[]}}

The first updated module is \texttt{MBRep[]}, which inputs the MB integral into \texttt{MBConicHulls.wl}. It now handles MB integrands with polygamma functions of arbitrary positive order. We have also added the option \texttt{PolygammaToGamma}, which lets the user either transform polygamma functions whose poles are split by the contours into gamma functions using~\eqref{PG2G}, or rewrite them as sums of polygamma functions whose poles are no longer split.

\mybox{
\textbf{\texttt{MBRep[PreFac,IntVar,MBVar,MBArg, Options[]]}}: This external module takes as input an MB integral of the form Eq.~\eqref{N_MB_3}.
\\\\
Below, we provide details about the input arguments for \texttt{MBRep}.
\begin{itemize}
    \item \texttt{PreFac}: Prefactor of the MB integral.
    \item \texttt{IntVar}: List of MB integration variables, such as $z_i$'s in Eq.~\eqref{N_MB_2}.
    \item \texttt{MBVar}: List of parameters, such as $x_i$'s in Eq.~\eqref{N_MB_2} each of which is raised to the power of $z_i$ in \texttt{IntVar}. {If the MB integral has no parameters, one can introduce some auxiliary scales as in Section~\ref{MBConicHulls.wl} and later set them to one when running the command \texttt{EvaluateSeries[]}}. However, the package also works if no scales are introduced.
     \item \texttt{MBArg}: List of two sublists of the form 
        {\tt \{\{Numerator\},\{Denominator\}\}}, where the elements of {\tt Numerator} include the arguments of numerator gamma functions. For numerator polygamma functions, it should be included in \texttt{Numerator} as a list of two elements, with the first element being the order and the second element being the argument of the polygamma function. {\tt Denonimator} are arguments of only the denominator gamma functions.
    \item Options:
    \begin{itemize}
    \item \texttt{Substitute}: This option accepts a list. Its default value is \texttt{\{\}}. The list provided substitutes the values of parameters appearing in the MB integrand.
    \item \texttt{TakeLimit}: This option accepts a list. Its default value is \texttt{
\{\}}. The list provided specifies the limits on parameters to be taken while transforming a straight to a non-straight MB.
    \item \texttt{CanonicalTransform}: This option accepts a boolean. Its default value is \texttt{True}. This option
    specifies whether to transform the input MB into a canonical MB of the form Eq.~\eqref{N_MB_3}.
    \item \texttt{PolygammaToGamma}: This option accepts a boolean. Its default value is \texttt{True}. This option
    specifies whether to rewrite numerator polygamma functions whose poles are split by a straight contour as derivatives of gamma functions in the input MB using Eq.~\eqref{PG2G}. If set to \texttt{False}, a transformation similar to Eq.~\eqref{straight2nonstraightPolyGamma} will be used, leading to more than one output MB.
    \end{itemize}
\end{itemize}
}

\subsection{\texttt{ResolveMB[]}}

The next updated module is \texttt{ResolveMB[]}, which applies the conic-hull method to identify the different sets of poles whose residues add up to form distinct series solutions. The input of this module is the output of \texttt{MBRep[]}. Therefore, if the user runs \texttt{MBRep[]} with \texttt{PolygammaToGamma $\to$ False}, then there may be more than one MB integral to be evaluated by \texttt{ResolveMB[]}. To handle this case, we have updated \texttt{ResolveMB[]} with an option \texttt{MBNumber} to select which MB integral is to be evaluated. Below, we provide its documentation.

\mybox{
\textbf{\texttt{ResolveMB[MBRepOut, Options[]]}}:
This external module takes as input the output returned by \texttt{MBRep[]}.
It then associates a set of conic hulls with the MB integral and finds the largest subsets of intersecting conic hulls to identify the sets of poles that contribute to different series solutions.
It also constructs the master conic hull and returns the characteristic list for the master series.
\\\\
Below, we provide details about the input arguments and options for \texttt{ResolveMB[]}.
\begin{itemize}
    \item \texttt{MBRepOut}: is the output of the \texttt{MBRep[]} function.
    \item Options:
    \begin{itemize}
    \item \texttt{MaxSolutions}: This option accepts a positive integer. Its default value is \texttt{Infinity}. Its value specifies the maximum number of series solutions of the MB integral that one wishes to evaluate. 
    \item \texttt{MasterSeries}: This option accepts a boolean. Its default value is \texttt{True}. Its value specifies whether to compute the master series for each of the series solutions found or not. 
    \item \texttt{PrintSolutions}: This option accepts a boolean. Its default value is \texttt{True}. Its value specifies whether to print the list of possible solutions along with their list of poles or not. Its default value is \texttt{True}.
    \item \texttt{MBNumber}: This option accepts a positive integer. Its default value is \texttt{1}. Its value
    specifies the number of the MB to be solved among the ones returned by \texttt{MBRep[]}.
    \end{itemize}
\end{itemize}
}

\subsection{\texttt{TriangulateMB[]}}

The next updated module is \texttt{TriangulateMB[]}, which applies the triangulation method to solve MB integrals and print sets of poles in a manner similar to \texttt{ResolveMB[]}. We have also added the new option \texttt{MBNumber} to this module to handle cases with more than one MB integral. Below, we provide its documentation.

\mybox{
\textbf{\texttt{TriangulateMB[MBRepOut,Options[]]}}: This external module takes as input the output returned by \texttt{MBRep[]}.
It then calls \texttt{TOPCOM}~\cite{Rambau2002} to find all the possible triangulations and prints the sets of poles for each possible series solution.
\\
Below, we provide details about the input arguments and options for \texttt{TriangulateMB[]}.
\begin{itemize}
    \item \texttt{MBRepOut}: the output of the \texttt{MBRep[]} function.
    \item Options:
    \begin{itemize}
        \item \texttt{MaxSolutions}: This option accepts a positive integer. Its default value is \texttt{Infinity}. Its value specifies the maximum number of series solutions of the MB integral that one wishes to evaluate. 
        \item \texttt{MasterSeries}: This option accepts a boolean. Its default value is \texttt{True}. Its value specifies whether to compute the master series for each of the series solutions found or not.
        \item \texttt{TopComParallel}: This option accepts a boolean. Its default value is \texttt{True}.
        Its value specifies whether to run \texttt{TOPCOM} in parallel or not. 
        \item \texttt{TopComPath}: This option accepts a string. Its default value is \texttt{"/usr/local/bin/"}. Its value specifies the path to the \texttt{TOPCOM} executables. 
        \item \texttt{PrintSolutions}: This option accepts a boolean. Its default value is \texttt{True}. Its value specifies whether to print the list of possible solutions along with their lists of poles or not.
        \item \texttt{ShortestOnly}: This option accepts a boolean. Its default value is \texttt{False}. Its value specifies whether to print only the solution with the lowest number of sets of poles or not.
        \item \texttt{MaxCardinality}: This option accepts a positive integer. Its default value is \texttt{None}. Its value specifies the maximum length (\textit{i.e.} number of sets of poles) of the solutions to be considered. 
        \item \texttt{Cardinality}: Its value specifies the length of the solutions that have to be considered. Its default value is \texttt{None}.
        \item \texttt{SolutionSummary}: This option accepts a boolean. Its default value is \texttt{False}.
        Its value specifies whether to print only a summary of solutions along with their cardinalities. 
        \item \texttt{QuickSolve}: This option accepts a boolean. Its default value is \texttt{False}.
        Its value specifies whether to find only the quickest possible solution. This is useful for higher-fold MB integrals. 
        \item \texttt{MBNumber}: This option accepts a positive integer. Its default value is \texttt{1}. Its value
        specifies the number of the MB integral to be solved among the ones returned by \texttt{MBRep[]}.
    \end{itemize}
\end{itemize}
}

\bibliographystyle{JHEP}
\bibliography{ref}
\end{document}